\documentclass[conference]{IEEEtran}

\usepackage{makeidx}  
\usepackage{graphicx}
\usepackage{amssymb}
\usepackage{algorithmic}
\usepackage{algorithm}

\usepackage{amsmath}
\usepackage{bm}
\usepackage{fancybox}

   \usepackage{setspace} 

\ifCLASSINFOpdf
\else
\fi

\usepackage{array}

\begin{document}

\title{An Optimized Signature Verification System for Vehicle  Ad hoc NETwork}

\author{\IEEEauthorblockN{Mohammad Saiful Islam Mamun~ and~ Atsuko Miyaji}
\IEEEauthorblockA{
Japan Advanced Institute of Science and Technology (JAIST)\\
Ishikawa, Japan.\\
\{mamun, miyaji\}@jaist.ac.jp}
}

\maketitle

\begin{abstract}
This paper\footnote{Research supported by Graduate Research Program (GRP), JAIST grants.} presents an efficient approach to an existing batch
verification system on Identity based group signature (IBGS)
which can be applied to any Mobile ad hoc network device including
Vehicle Ad hoc Networks (VANET). We propose an optimized way to
batch signatures in order to get maximum throughput from a
device in runtime environment. In addition, we minimize the number
of pairing computations in batch verification proposed by B. Qin et al. for
 large scale VANET. We introduce a batch scheduling algorithm for batch verification targeting further minimization the batch computation time.

\end{abstract}

\IEEEpeerreviewmaketitle

\section{Introduction}
VANET is recently a very popular
direction for research targeted to proper safety and efficiency
in transport management systems.High mobility, high speed of vehicles, fast topology changes, and sheer scale are some characteristics that establish VANET as an intensive research point different from other types of ad hoc networks. Recently some security hardwares are introduced to VANET
vehicles which makes it feasible to implement robust cryptographic tools \cite{2}. For example, Event Data Recorders (EDR) which record all received messages s.t., position data, speed data, acceleration data, time etc.; Tamper Proof Devices (TPD) which provide the ability of processing, signing and verifying
messages, protect hardware from tampering by a set of sensors, possess its own battery and clock, etc.  
Vehicle manufacturers along with telecommunication
industries are encouraged to equip each car with On Board Units
(OBUs) that allow vehicles to communicate with each other, as
well as to supply Road Side Units (RSUs). The OBUs together with
RSUs of a vehicular ad hoc network (VANET) help to broadcast
messages to other vehicles or transport system terminals in
their range. It is a serious threat to privacy if
attackers can track the drivers' geographical position, identity,
or his driving pattern by monitoring vehicular
communication \cite{4}. Alternatively, there is no way to identify
rogue vehicles, if the network is completely anonymous.\\

~Conventional signature verification mechanisms might not be sufficient to satisfy the stringent
time requirement in VANET. For example, in a VANET environment hundreds of vehicles are
connected to one another at any time instance and each of them may send
messages every 100-300 \textit {ms} within a period of 10 \textit {s} travel time.
Therefore, a vehicle needs to verify hundreds of message
signatures per second \cite{6}. Although only one signature is received by Dedicated Short Range Communication (DSRC) transmission at a time, a large number of signatures are buffered at receiving
station. According to \cite{7}, consumption period of a DSRS transmission is too shorter than that of verifying a signature. For this reason, verifying a huge number of messages one after one is impractical and may cause bottlenecks at TPD in the vehicle. Moreover, some signatures might arrive from emergency vehicles like ambulance, petrol police, security van etc. that incur urgent response (short due time) from the receivers. As a consequence, many messages coming from other vehicles may be discarded due to time constraints or proper scheduling.\\ 

~In order to deal with all the challenges mentioned above [2,4-7], a number of batch verification mechanisms have been proposed [1,10,11]. In this paper, we consider the batch verification described in \cite{main} that was designed for large scale VANET. They introduced several features like (i) an Identity based Group signature with easy-to-manage smaller groups, (ii) Revoking the anonymity of signatures by Open Authority in case of disputed or suspected forged messages, (iii) A selfish verification mechanism to speed up signature processing. We use the same group signature mechanism in \cite{signature} as \cite{main} used to provide anonymity and traceability. To alleviate the burden of signature verification, we further improve batch verification described in \cite{main}. For example, we reduce the calculation of pairing from 11 to 3 per signature. In addition, we introduce a batch scheduling algorithm to reduce the time consumption of batch verification at a greater extent. Besides that we include revocation procedure if the message is doubtable to be corrupted.\\ 

~The remainder of the paper is organized as follows. In section II, background and preliminary knowledge related to the proposed research are given, including network model, basics of batch verification, and security requirements. In section III, IBGS scheme used in this paper is described in detail. Section IV discusses the proposed batch verification scheme including batch size and signature scheduling algorithm and doubtable message verification technique. In section V, security analysis and performance evaluation are presented in brief. Finally, section VI concludes the paper.
\section {Preliminaries}
\subsection {Assumptions.}
We assume that Trusted Escrow  Authority (TEA) is fully trusted. Other entities e.g., Vehicles, Group Managers (GM), and Opening Authority will need to register before operation. Registration of the vehicle may include the vehicle\rq{s} number plate, identity, and driver information to recognize the vehicle and its owner uniquely. However, GMs are well trusted since Access Networks are usually set up in an open environment which is vulnerable to security breaches. That is why, TEA should check the security parameters of the GMs periodically.

 \subsection {Network Model.}
 Our network model is shown in Fig. 1., where On-Road unit refers to the units on the roadside e.g. vehicles, Access Network Manager. Groups can be formed in many ways. For example, by \textbf {region}  or area where the vehicles are registered e.g., New York city, Tokyo city etc.; by \textbf {social spot} e.g., shopping mall, official zone, military zone, educational institutions etc.; by \textbf {category} of the vehicles e.g., personal cars, ambulance, police cars, fire trucks etc. These groupings help in applying policy to manage the vehicles intelligently. This is a hierarchical network, from TEA to vehicles. All the vehicles are attached, accessed and managed by Group Managers and Group Managers are subsequently attached to TEA. Alternatively, Traffic Security Division is attached with both Off-Road and On-Road units simultaneously.   TEA is responsible for issuing key, revocation. Each device need to be equipped with On Board Unit (OBU) consisting of a temper proof device(TPD), Global Positioning System (GPS) an Event Data Recorder(EDR). Usually a Road Side Unit has DSRC with a vehicle\rq{s} OBU. All communications such as Vehicle-to-Vehicle and Vehicle-to-AccessNetwork should be time synchronized. 
   \graphicspath{{\\C:\Users\Mamun\Desktop\Accepted paper\IEEEtran (1)}} 
\begin{figure}[!t]
\centering
\includegraphics[width=0.5\textwidth]{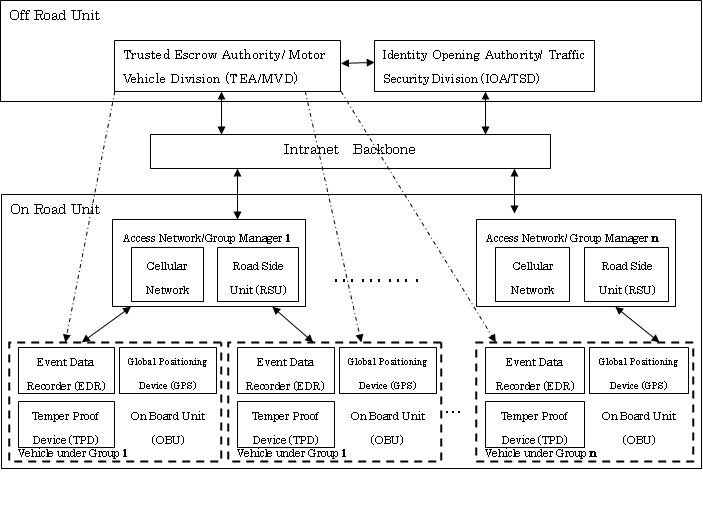}
\caption{VANET Network Model}
\label{fig: VANET_Network_Model}
\end{figure}

   \subsection {Batch verification.}

\subsubsection {Small Exponent test} M. Bellare et al. \cite{mihir_batch} gave the first idea about fast batch verification on digital signatures. They proposed Small Exponent test for batch verification.

    \begin{itemize}
      \item Choose $\delta_1....\delta_n \in \{0,1\}^l$
      \item Compute $ a = \sum_{j=1}^{n} a_j \delta_j ~mod ~q $ and $ y= \prod_{j=1}^{n} y_j^{\delta_j}$. where $a_j, y_j \in \mathbb{Z}_{q} $ 
            \item Check whether it holds: $g^a=y$. if yes accept, else reject.

\end{itemize}
We can expect low error from the Small Exponent Test if we choose $\delta_1....\delta_n$ from a large domain.

\subsubsection {Basic Batch Techniques} A.L. Ferrara et al. \cite {ferrara_batch} proposed 3 techniques to develop an efficient batch verifier for bilinear equation which is applied in our scheme. Here are their techniques in brief:
\begin{itemize}
\item \textbf {Technique 1.} Sigma-protocols, usually called $\sum$-protocols, have three move structures: commitment, challenge and response. This is one of the implementable protocols of \textit {Proof of knowledge}. These three steps degrade the verification mechanism worse. Ferrara et. al. suggest to reduce it as (commitment,response) policy to achieve much more verifiable equations. For pairing, they propose two sub-steps:
  \begin{itemize}
\item \textit{Check membership:} Only elements that an adversary could attack need to be checked. Public parameters need not be checked, or it can be checked once. 
\item \textit {Small Exponent Test:} Perform the test to combine all the equation into one.
 \end{itemize}
  \item \textbf {Technique 2.} Move the exponent into the pairing, for example, Replace \textbf {$~~e(g_i,h_i)^{\delta^i}$} with \textbf {$e(g_i^{\delta^i},h_i)$}. It speeds up the exponentiation process.
   
 \item \textbf {Technique 3.} If two pairing with a common elements appear. It will reduce $n$ pairing to $1$.For example, replace $\prod_{i=1}^{n} e(g_i^{\delta^i},h)$ with  $e(\prod_{j=1}^{n} g_i^{\delta^i},h)$  
\end{itemize}

\subsection {System Security Requirement.}
VANET is concerned with  public safety and vehicle privacy. An attacker who usually observes and analyzes the traffic in the network is called passive or external attacker. However, a compromised vehicle or device,  called an internal attacker, can be more influential.  An IBGS scheme in a VANET environment needs to ensure correctness, anonymity, traceability, nonframeability, revocability,liability and scalability as security and system requirement. We will give a brief description of these properties.

          \begin{itemize}
\item \textbf {Anonymity.} A signature scheme is anonymous if no PPT adversary can identify the message originator by monitoring the communication between vehicles or access points as a passive attacker.  According to privacy requirements, vehicle and owner information needs to be secret. Anonymity mechanism ensures that any identification related to vehicle will not be advertised or be traceable except in some special situation by some designated party.

\item \textbf {Revocability.} If the vehicle doubts a received message with a valid signature then it should be able to send the message to the verifier to judge identification of the message sender and hence revoke a malicious signature. Unfortunately this is a threat to privacy, but for the sake of public safety, personal privacy needs to be sacrificed.  If anonymity is realized without any revocability mechanism, an inside attacker can anonymously broadcast forged messages to  other vehicles, which may seriously degrade public safety and system reliability as well.

\item \textbf  {Liability.} Liability demands that the sender of the message should be responsible for the message generated. Authentication, integrity and nonrepudiation must be supported  in the network to ensure liability.

\item \textbf  {Traceability.} A scheme is traceable if any polynomial-time attacker has only a negligible probability
of producing a valid group signature such that the output is not the identity of the group signature originator.
\item \textbf  {Correctness.}The signature scheme is correct if it has verification and opening correctness.
\item \textbf  {Scalability.} In a densely populated metropolitan area, every day hundreds of vehicles are registered, stolen, and loose secret key. It entails extra job of message signature, verification, judging etc. while preserving all the security requirements. Therefore,  it is essential for the network to be scalable for smooth management.

\item \textbf  { Non-frameability.} It requires that the attacker be unable to created a judge-accepted proof that an honest user produced a certain vald signature unless this user really did produce this signature. 

          \end{itemize}

\section{VANET based IBGS}

An ID based group signature scheme \cite {signature}, where there are group managers, group
members, and the Open Authority, provides a full traceability, a full anonymity and non-frameability as needed for VANET
application. Thats why, \cite {main} chooses \cite {signature} for large scale VANET. We summarize \cite {signature} as follows:

\subsection {Setup$( 1^l)$.} Let a security parameter $(1^l)$ ,$p$ is a prime and finite cyclic groups $\mathbb{G}_1= \langle A \rangle$ and $\mathbb{G}_2= \langle B \rangle$, where there exist a computable isomorphism $ \Psi = \mathbb{G}_1 \to \mathbb{G}_2$ and a non-degenerate bilinear pairing $ e: \mathbb{G}_1 \times \mathbb{G}_2 \to \mathbb{G}_3$.
For initial setup \textbf {TEA} will do the followings: 

	\begin{enumerate}
            \item Set $\Re = ( p,\mathbb{G}_1, \mathbb{G}_2, \mathbb{G}_3,A,B,e ) $ 
            \item Choose $A_1, A_2,A_3, A_4, A_5 \in \mathbb{G}_1$
            \item Define cryptographic function
            \begin{itemize}
               \item $ H_V : {\{0,1\}}^* \to \mathbb{G}_1 $ for Vehicles
               \item $ H_O : {\{0,1\}}^* \to \mathbb{G}_1 $ for TSD 
               \item $ H_R : {\{0,1\}}^* \to \mathbb{Z}_{p}^{*} $ for GM 
               \item $ H : {\{0,1\}}^* \to \mathbb{Z}_{p}^{*} $ for vehicles to compute or, to verify challenges. 
            \end{itemize}
            \item Generate \textit {private} Key $(x_T) \in \mathbb{Z}_{p}^{*} $ and \\ \textit {public} key $ (K_T) = B^{x_T} \in \mathbb{G}_2 $
               
             \item Finally, produce the system\rq{}s public parameter:\\  $\textit param   =( \Re, A_1, A_2,A_3, A_4, A_5, H_V, H_O, H_R, H,K_T) $
           	\end {enumerate}
           	
 \subsection  {Key Generation.}TEA generates private keys for all the entities in the system including vehicles, GMs, and TSD with the identities they provide.
               \begin{itemize}
            \item \textit {GM:} With the identity of GM $ID_R$, TEA generates private key $x_R= r + H_R (C||ID_R)x_T~~ mod~~ p$  where $r \in \mathbb{Z}_{p}^{*}$ and $ C = B^r $. TEA issues $(C,x_R)$ to each Group Manager. It is to mention that TEA provide the additional string $C$ to all the GM to register with corresponding vehicles. 
             \item \textit {TSD:} Private key $x_O = H_O (ID_O)^{x_T}$ where $ID_O$ is the identity of TSD.
             \item \textit {Vehicle:} Private key   $x_V = H_V (ID_V)^{x_T}$ where $ID_V$ is the identity of a vehicle.
             \end{itemize}
          	
Vehicles need to complete registration with their corresponding GM. GM firstly runs a proof of knowledge of vehicular private key $x_V$ for its Identification without any information leakage. Additionally GM sends the additional string $C$ as it got from TEA. It also manages a registration table or database to entry the vehicle as a group member. 
 \subsection{Group Join Protocol.} TEA is the only trusted authority. Before joining the group, a vehicle has its ID and secret key generated from TEA. To get a membership certificate a vehicle needs to perform a protocol with the certificate issuer called GM as shown in \textbf {Fig. 2.}
\begin{figure*}
\normalsize
\begin{center}
\begin{tabular}{lcl}
\hline
\hspace{-0.30cm} \bfseries Vehicle $(x_V)$ & &\hspace{-0.30cm} \bfseries GM $(C,x_R)$\\
\hline
& & \hspace{-0.30cm}Run Proof of Knowledge for $x_V$\cite{Decryp}\\
& & \hspace{-0.30cm}Compute $D = {A_5/  H_V (ID_V)}^{1/(t+x_R)} $~$ t \in \mathbb{Z}_{p}^{*} $\\
& & \hspace{-0.30cm}Compute $W =  e( H_V (ID_V) ,B)$\\
& & \hspace{-0.30cm}Record $(ID_V, D,t,W)$ for future use. \\
\hspace {-0.30cm}& $\underleftarrow {(D, t, C)}$ & \\
\hspace {-0.30cm}Check validity of $C$ \cite {Bellare_IDS} and accept if & &\\
\hspace {-0.30cm}$ e(A_5, B) = e(D,B)^t e(D,S) e( H_V (ID_V) ,B)$ & &\\
\hspace {-0.30cm}holds for $S = C {K_T}^{H_R(C||ID_R) }$ & &\\
\hline
\end{tabular}
\end{center}
\caption{Group Joining Protocol}
\vspace*{4pt}
\end{figure*}
At the end of the protocol, a vehicle becomes a member of the group and obtains a membership certificate $(D, t, C)$ as a \textit {Group Signing Key}. GM computes $W$ which is stored in the database for future use. In case of irregular behavior, $W$ will help TSD to reveal the identification of a vehicle.

\subsection  {Signing and Authentication.}
 A registered vehicle under a group having a secret key $x_V$ and  \textit {Group Signing Key}$ (D,t,C)$ can anonymously generate a signature $\Upsilon$ on a message $M$. At the same time, it allows TSD, or Open authority to open the signature if needed. Detailed descriptions are as follows:  \\
 \begin {enumerate}

       \item Choose $s_1 \in \mathbb{Z}_{p}^{*}$ and set $(\Gamma_0,\Gamma_1,\Gamma_2,\Gamma_3,\Gamma_5, s_2 ) = \\(A^{s_1}, x_VA_1^{s_1}, H_V (ID_V)A_2^{s_1}, DA_3^{s_1},\Gamma_3^t A_4^{s_1} , ts_1~mod ~p)$
  
       \item Choose $d \in \mathbb{Z}_{p}^{*}$ and  set $(v_1,V_2)=\\ \Big (e(H_V (ID_V),B)  e(H_O (ID_O),K_T)^d\textbf {,}~ B^d\Big)$      

 \item Select randomly $(r_1,r_2,r_3,r_4) \in \mathbb{Z}_{p}^{*}$ and \\$(R_1,R_2,R_3) \in \mathbb{G}_1 $ and compute
         \begin {itemize}  
            \item $(\beta_0,\beta_1,\beta_2,\beta_3,\beta_5,\beta_7) =\\ \Big(A^{r_1}, R_1A_1^{r_1},R_2A_2^{r_1},R_3A_3^{r_1}, \Gamma_3^{r^3}A_4^{r_1}, B^{r_4}\Big) $
            \item  $(\beta_4,\beta_6,\beta_8) = \Big( [e(A_1,B)^{-1} e(A_2, K_T)]^{r_1},
            e(A_3,B)^{r_2}\\ {[e(A_3,S) e(A_2A_4, B)]}^{r_1},e(H_O (ID_O),K_T)^{r_4} e(A_2,B)^{-r_1} \Big)  $
             
         \end {itemize}   
         \item Compute
         $f = H \Big((\Gamma_0,\Gamma_1,\Gamma_2,\Gamma_3,\Gamma_5)||C||v_1||V_2||M ||\\~~~~~~~~~~(\beta_0,\beta_1,\beta_2,\beta_3,\beta_4,\beta_5,\beta_6,\beta_7,\beta_8)\Big)$\\
         \item Compute
          \begin {itemize}   
          \item$(z_0,z_1,z_2,z_3) = \Big(r_1 - fs_1 ~mod~ p,\\r_3-ft ~mod ~p, r_2-fs_2 ~mod ~p,
           r_4-fd~mod~p\Big)$
          \item$(Z_1,Z_2,Z_3) = ( R_1x_V^{-f}, R_2H_V(ID_V)^{-f}, R_3D^{-f})$
          \end {itemize}   
          \item Signature for batch verification \\$\Upsilon = \Big(\Gamma_0,\Gamma_1,\Gamma_2,\Gamma_3,\Gamma_5)||(z_0,z_1,z_2,z_3,Z_1,Z_2,Z_3) ||f||\\~~~~~C||v_1||V_2 || (\beta_0,\beta_1,\beta_2,\beta_3,\beta_4,\beta_5,\beta_6,\beta_7,\beta_8)|| S\Big)$
                   
 \end {enumerate}
\subsection {Individual Message Verification.} After getting a signature $\Upsilon$, a vehicle verifies the signature as follows:
\begin {itemize}

\item Set $(\Gamma_4\textbf {,}\Gamma_6\textbf {,}\Gamma_8) = (e(\Gamma_1,B)^{-1} e(\Gamma_2, K_T)\textbf {,}~ e(A_5,B)^{-1}\\e(\Gamma_3,S) e(\Gamma_2\Gamma_5, B) \textbf {,}~V_1e(\Gamma_2,B)^{-1} )$
\item Compute $S = C {K_T}^{H_R(C||ID_R) }$
\item Compute $(\beta_0,\beta_1,\beta_2,\beta_3,\beta_5,\beta_7)=\\ \Big (A^{z_0}\Gamma_0^f\textbf {,} 
Z_1 A_1^{z_0}\Gamma_1^f\textbf {,}
Z_2A_2^{z_0}\Gamma_2^f\textbf {,}
 Z_3A_3^{z_0}\Gamma_3^f\textbf {,}  
\Gamma_3^{z_4}A_4^{z_0}\Gamma_5^f\textbf {,}
B^{z_6}V_2^f \Big)$

\item Compute$(\beta_4,\beta_6,\beta_8)=
\Big( [e(A_1,B)^{-1}e(A_2, K_T)]^{z_0} \Gamma_4^f\textbf {,}\\e(A_3,B)^{z_5}{[e(A_3,S) e(A_2A_4, B)]}^{z_0}\Gamma_6^f\textbf {,}\\e(H_O (ID_O),K_T)^{z_6} ~e(A_2,B)^{-z_0}\Gamma_8^f \Big )$ 

\item Check $f =  H \Big((\Gamma_0,\Gamma_1,\Gamma_2,\Gamma_3,\Gamma_5)||C||v_1||V_2||M||\\(\beta_0,\beta_1,\beta_2,\beta_3,\beta_4,\beta_5,\beta_6,\beta_7,\beta_8) \Big)$ 
 \end{itemize}
If the hash check is successful, then the message will be accepted; otherwise rejected.

\section {The Proposal}

We start with VANET based IBGS scheme described in section III and reconsider the signature generation in \cite {main} to remove some redundant parts\footnote{Size of the signature in \cite {main} is larger than \cite {signature}.}. Then, we modify the verification algorithm to apply an efficient batch verification. Finally, we propose a batch scheduling algorithm to gear up the verification in a runtime environment. 
\subsection {Modified Signature.} We follow the same signature in section III, but modify the final part of signature. Here we describe the only the modified portion. A signer who possesses \textit {group signing key} can anonymously generate a signature on a message $M$.
\begin{itemize}
\item  Compute  $(\beta_4,\beta_6,\beta_8)$ as section III.

\item Modified signature will be:\\ $\Upsilon^{\prime}=(\Gamma_0,\Gamma_1,\Gamma_2,\Gamma_3,\Gamma_5) || (z_0,z_1,z_2,z_3,Z_1,Z_2,Z_3) ||\\(\beta_4,\beta_6,\beta_8)||f||C||v_1||V_2$
   
\end{itemize}

\subsection {Modified Individual Verification.} We have observed that $(\beta_4,\beta_6,\beta_8)$ is the most expensive part of the verification.
In section III, we gave the verification mechanism of \cite {main} which is modified here according to the above signature $\Upsilon^\prime$ to reduce the cost of individual verification. To verify signature $\Upsilon^\prime$ of a message $M$ one can perform the following:

\begin{itemize}
\item Compute $S = C {K_T}^{H_R(C||ID_R) }$

\item  Compute $(\beta_0,\beta_1,\beta_2,\beta_3,\beta_5,\beta_7) = \\ ( A^{z_0}\Gamma_0^f, Z_1 A_1^{z_0}\Gamma_1^f, Z_2A_2^{z_0}\Gamma_2^f, Z_3A_3^{z_0}\Gamma_3^f, \Gamma_3^{z_4}A_4^{z_0}\Gamma_5^f, B^{z_6}V_2^f )$

\item Check $f =  H \Big ((\Gamma_0,\Gamma_1,\Gamma_2,\Gamma_3,\Gamma_5)||C||v_1||V_2||M||\\~~~~~(\beta_0,\beta_1,\beta_2,\beta_3,\beta_4,\beta_5,\beta_6,\beta_7,\beta_8) \Big)$ 

\item Verify  $(\beta_4,\beta_6,\beta_8) =\\ \Big ([e(A_1,B)^{-1} e(A_2, K_T)]^{z_0}  [e(\Gamma_1,B)^{-1} e(\Gamma_2, K_T)]^f,\\e(A_3,B)^{z_2}[e(A_3,S) e(A_2A_4, B)]^{z_0} [e(A_5,B)^{-1}e(\Gamma_3,S)\\ e(\Gamma_2\Gamma_5, B)]^f,
 e(H_O (ID_O),K_T)^{z_6} e(A_2,B)^{-z_0}[v_1e(\Gamma_2,B)^{-1}]^f \Big)$ 
\end{itemize}
Compared with the verification mechanism of \cite{main}, we noticed that  $( \Gamma_4,\Gamma_6,\Gamma_8), S$ and $(\beta_0,\beta_1,\beta_2,\beta_3,\beta_5,\beta_7) $ are not necessary to be included in a signature, and thus, signature size is reduced.

\subsection {Modified Batch Verification.} The scheme in \cite {main} exploits the techniques of \cite {ferrara_batch} keeping in mind that a multi-base exponentiation (pairing) takes a similar time as a single-base exponentiation. Let a VANET device receives $n$ message-signature pair $ (M_i,  {\Upsilon^\prime}_i)$ where ${\Upsilon_i}^{\prime} = (\Gamma_{0,i},\Gamma_{1,i},\Gamma_{2,i},\Gamma_{3,i},\Gamma_{5,i}) ||(\beta_{4,i},\beta_{6,i},\beta_{8,i})||f_i||C_i||v_{1,i}||V_{2,i}||\\ (z_{0,i}, z_{1,i},z_{2,i},z_{3,i},Z_{1,i},Z_{2,i},Z_{3,i})$.\\

   \begin{itemize}
   
     \item Compute  $S = C {K_T}^{H_R(C||ID_R) }$ once, because all vehicles share the parameters $C, ID_R, K_T$ [\textit {Technique 1}, in section 2.]
          
     \item  For all $i = 1,....,n$ compute the non-pairing equations:$(\beta_{0,i},\beta_{1,i},\beta_{2,i},\beta_{3,i},\beta_{5,i},\beta_{7,i}) = \\ ( A^{z_0}\Gamma_{0,i}^{f_i},
         Z_{1,i} A_1^{z_{0,i}}\Gamma_{1,i}^{f_i},
          Z_{2,i}A_2^{z_{0,i}}\Gamma_{2,i}^{f_i},
           Z_{3,i}A_3^{z_{0,i}}\Gamma_{3,i}^{f_i}, \Gamma_{3,i}^{z_{4,i}}\\A_4^{z_{0,i}}\Gamma_{5,i}^{f_i},
 B^{z_{6,i}}V_2^{f_i} )$

 \item For each $i = 1,........,n$ check the following: \\
 $f_i = H \Big((\Gamma_{0,i},\Gamma_{1,i},\Gamma_{2,i},\Gamma_{3,i},\Gamma_{5,i})||C_i||v_{1,i}||V_{2,i}||M_i||\\(\beta_{0,i},\beta_{1,i},\beta_{2,i},\beta_{3,i},\beta_{5,i},\beta_{7,i})\Big )$
 
  \item Before starting a batch verification, we can simplify the equation as follows:
 
   $  \beta_4  = [e(A_1,B)^{-1} e(A_2, K_T)]^{z_0}  [e(\Gamma_1,B)^{-1} e(\Gamma_2, K_T)]^f \\
  ~~~~~  = e({A_1}^{-z_0} , B) e({A_2}^{z_0}, K_T) e({\Gamma_1}^{-f},B) e({\Gamma_2}^f, K_T)\\~~~~= e({A_1}^{-z_0} \Gamma_1^{-f}, B)~~ e({A_2}^{z_0} {\Gamma_2}^f, K_T)$

$\beta_6 = e (A_3,B)^{z_5}[e(A_3,S) e(A_2A_4, B)]^{z_0} [e(A_5,B)^{-1}\\ ~~~~~~~~
e(\Gamma_3,S)e(\Gamma_2\Gamma_5, B)]^f \\
~~~~~=e ({A_3}^{z_5},B) e({A_3}^{z_0},S) e({A_2}^{z_0}{A_4}^{z_0}, B) e({A_5}^{-f},B) \\~~~~~~~~e({\Gamma_3}^f,S) e({\Gamma_2}^f {\Gamma_5}^f, B)\\
~~~~~= e ({A_3}^{z_5} {A_2}^{z_0}{A_4}^{z_0}~{A_5}^{-f}{\Gamma_2}^f {\Gamma_5}^f,B) ~~e({A_3}^{z_0}  {\Gamma_3}^f,S)$

$\beta_8 =  e(H_O (ID_O),K_T)^{z_6} ~~e(A_2,B)^{-z_0}[v_1~~e(\Gamma_2,B)^{-1}]^f \\
~~~~~= e({[H_O (ID_O)]}^{z_6},K_T) ~~e{(A_2}^{-z_0},B) ({v_1}^f)~~ e({\Gamma_2}^{-f},B)\\
~~~~~=({v_1}^f) ~~e({A_2}^{-z_0} {\Gamma_2}^{-f},B)~~ e({[H_O (ID_O)]}^{z_6},K_T) $

 \item Let $ \xi_b= {A_1}^{-z_0} \Gamma_1^{-f},\\
 \xi_k= {A_2}^{z_0} {\Gamma_2}^f,\\
  \zeta_b= {A_3}^{z_5} {A_2}^{z_0}{A_4}^{z_0}{A_5}^{-f} {\Gamma_2}^f {\Gamma_5}^f,\\ 
  \zeta_s = {A_3}^{z_0}  {\Gamma_3}^f,\\
    \chi_b ={A_2}^{-z_0} {\Gamma_2}^{-f},\\
  \chi_k= {[H_O (ID_O)]}^{z_6}$

     \item   Hence       $(\beta_4\beta_6\beta_8) \\
      = (e(\xi_b,B) e(\xi_k,  K_T) e( \zeta_b,B) e( \zeta_s,S) e(\chi_b,B)\\~~~~~e(\chi_k, K_T) ({v_1}^f))\\
      =  (e(\xi_b\zeta_b,B) ~~e(\xi_k\chi_k,K_T)~~ e( \zeta_s ,S)({v_1}^f))$

        \item Applying \textit {Technique 1,3} in section II, choose the random vector $(\delta_1,....,\delta_l )$ where $\delta_i \in \mathbb{Z}_{p}$; and check the following pairing equations $\forall i = 1,........,n$:\\ 

  $ \prod_{i=1}^{n}(\beta_{4,i}\beta_{6,i}\beta_{8,i})^{\delta_i} 
      \\=  e( \prod_{i=1}^{n} {\delta_i} \xi_{b,i}\zeta_{b,i},B)~~ e(\prod_{i=1}^{n}{\delta_i} \xi_{k,i}\chi_{k,i},K_T) \\~~~~e(\prod_{i=1}^{n} {\delta_i} \zeta_{s,i} ,S)~~(\prod_{i=1}^{n}{\delta_i} {v_{1,i}}^{f_i})$ \\      

For simplicity let:~~\\
      $\mathbf{M_i}= (\beta_{4,i} \beta_{6,i}\beta_{8,i})^{\delta_i},~
      \mathbf{B_i} =   {\delta_i} \xi_{b,i}\zeta_{b,i}, \\ \mathbf{Q_i} = {\delta_i} \zeta_{s,i}, ~
      \mathbf{K_i}  = {\delta_i} \xi_{k,i}\chi_{k,i} ~$ and $~
          \mathbf{\nu_i} = {\delta_i}{v_{1,i}}^{f_i}$\\

Hence, $  \prod_{i=1}^{n} \mathbf{M_i}  =  e( \prod_{i=1}^{n} \mathbf{B_i},B)  ~~  e(\prod_{i=1}^{n}\mathbf{K_i} ,K_T) ~~ \\e(\prod_{i=1}^{n} \mathbf{Q_i}  ,S)   \prod_{i=1}^{n} \mathbf{\nu_i}  $\\  
        
If the above equation is satisfied then verification is successful; otherwise not. 
\end {itemize}

 \subsection {Doubtable Message Verification.} Consider the verifying entity receives a message which has a valid signature but the message is doubtable to be forged. It might happen if the signer's secret key is compromised. If the message is found to be fraudulent then usually the certificate of the compromised signer is revoked and the revocation list updated. In \cite {main} the authors suggested \textit {a tag} specifying the lifetime of the GM\rq{}s public key. TSD has the right to open the encryption in the signature. To trace the actual signer of a given signature $\Upsilon$, TSD does the following:

\begin{itemize}
\item	The verifier submits the message M with its corresponding signature $\Upsilon$ to TSD which computes: $v_1/e(x_O,V_2) = e(H_V(ID_V),b)= \upsilon $

\item TSD compare $\upsilon$ with the entry in registration table. If no entry is found, output $\bot$; else computes proof of knowledge  $\omega$ s.t. $ e(x_O,V_2)= v_1/\upsilon$ to justify.

\begin {enumerate}
    \item Select (${s_0}^\prime,  {r_0}^\prime,{r_1}^\prime) \in \mathbb{Z}_{p}^{*}$  and compute :
    $(\Gamma_0^\prime, \Gamma_1^\prime, \Gamma_2^\prime) = (x_OA^{{s_0}^\prime}, e(A,V_2)^{{s_0}^\prime}, e(A,B)^{{s_0}^\prime} )$
         
          \item  $(\beta_0^\prime,\beta_1^\prime,\beta_2^\prime) =(  H_O(ID_O)^{{r_1}^\prime}A^{{r_0}^\prime},e( A,V_2)^{{r_0}^\prime},\\ e(A,B)^{{r_0}^\prime} )$
           \item Compute $f^\prime= H ((\Gamma_0^\prime, \Gamma_1^\prime, \Gamma_2^\prime)|| (\beta_0^\prime,\beta_1^\prime,\beta_2^\prime) ||v_1||V_2,\upsilon)$
          \item $ ({z_0}^\prime,{z_1}^\prime) = ({r_0}^\prime  - f^\prime {s_0}^\prime), H_O(ID_O)^{{r_1}^\prime}{x_O}^{f^\prime})$
  \item Outputs the proof: 
   $ \omega = (\Gamma_0^\prime||f^\prime||({z_0}^\prime,{z_1}^\prime) ) $ 
\end {enumerate}
\end{itemize}
\textit {Justification:} GM or, corresponding vehicle now can check the validity of the proof $\omega$ whether ID is the real signer of corresponding signature $\Upsilon$ for message $M$.
 \begin{itemize}
\item Compute:
$ (\upsilon,\upsilon^\prime) = (e(H_V(ID_V), B), v_1/\upsilon ) $ and
\item   $ (\Gamma_1^\prime,  \Gamma_2^\prime)  = (e ( \Gamma_0^\prime ,V_2 ) /{M^\prime}, e( \Gamma_0^\prime ,B ) e(H_O(ID_O),K_T)$
  \item $  (\beta_0^\prime,\beta_1^\prime,\beta_2^\prime) = ( {z_1}^\prime){\Gamma_0^\prime}^{f^\prime}A^{{z_0}^\prime},    e(A,V_2)^{{z_0}^\prime}{\Gamma_1^\prime}^{f^\prime},\\ e(A,B)^{{z_0}^\prime}{\Gamma_2^\prime}^{f^\prime}  )$
  \item Compare:
$f^\prime= H ((\Gamma_0^\prime, \Gamma_1^\prime, \Gamma_2^\prime)|| (\beta_0^\prime,\beta_1^\prime,\beta_2^\prime) ||v_1||V_2,\upsilon)$   
\end{itemize}	      	
If the above equation holds, the verifier will be assured about the message signer\rq{'}s ID as a real signer, hence justified.

\subsection {Batch Scheduling.}

In a VANET environment, signatures arrive sequentially with respect to time. OBU usually processes the signature First Come First Serve (FCFS) basis. But we cannot apply FCFS for $n$ number of signatures, because sometimes prioritized emergency data like message from \textit {fire service vehicle}, \textit {ambulance} etc. might be arrived with lower \textit {deadline} and/or higher \textit {weight} value.
That is why, it is practical to process the signatures according to their priority. In the proposed algorithm, we consider \textit {due time} of each signatures by which it can be scheduled to get faster response  and we also introduce the parameter \textit {weight} for each signature. By default, \textit {weight} could be a fixed value for example- \lq{}1\rq{}. But in special cases, it might have some different value which results providing higher priority to certain vehicles e.g., vehicle of prime minister, doctors etc. In addition, we can make different groups of vehicles with different \textit {weight} as a parameter e.g., public service buses, personal cars etc. Besides these,   
it is obvious that if we increase the \textit {batch size} $n$, it will give us more optimized computation time for $n$ signatures but also increases \textit {completion time} $C_i$ of individual signature. Therefore, there is a trade-off between the \textit {batch size} and {completion time}. Choosing the value of $n$ is not easy task. For example, the value of $n$ during rush hour at downtown could be greater than off-peak traffic times, or the same at uptown. For the \textit {batch verification} of size $n$, a vehicle needs waiting for $n$ signatures to arrive to commence verification. However, we can utilize idle time (waiting time) with partial computations of incoming signatures as it arrives while batch verification continues. We carefully observe that the right side of each batch verification pairing:$~~e(\prod_{i=1}^{n} \mathbf{B_i},B), ~e(\prod_{i=1}^{n}\mathbf{K_i},K_T),~$and$ ~e(  \prod_{i=1}^{n}\mathbf{Q_i} ,S)~~$ is common which contribute to partial computation of left part for consecutive signatures.
So, we will discuss the scheduling of signatures for partial computation and hence devise an algorithm to find an optimal number $n$ of signatures to be batched at a time. That\rq{}s why, We split the $i^{th}$ signature\rq{}s pairing verification  into 2 parts:\\

 \begin {itemize}
   \item \textbf {Verification Part 1:} Compute $   \mathbf{M_i},~~ \mathbf{B_i},~~ \mathbf{K_i},~~ \mathbf{Q_i}, ~$and$ ~ \mathbf{\nu_i} $ of corresponding signature with single machine scheduling policy as it arrives.
   \item \textbf {Verification Part 2:} Compute and verify: $ \prod_{i=1}^{n} \mathbf{M_i}  =\\  e( \prod_{i=1}^{n} \mathbf{B_i},B)  ~~  e(\prod_{i=1}^{n}\mathbf{K_i} ,K_T) ~~ e(\prod_{i=1}^{n} \mathbf{Q_i}  ,S)   \prod_{i=1}^{n} \mathbf{\nu_i}  $\\
   
   \end {itemize}

At first, we consider \textbf {Single Machine Scheduling Problem} with Release Times and Identical Processing Times for batch scheduling \cite {Batch Scheduling}; where the release times, due dates, and sequence-dependent set-up times of jobs are taken into consideration. The objective is to find a feasible set of jobs which are to be completed before or at their deadlines. Secondly, an algorithm is devised to find the optimal number of signatures to be processed in a batch.

Suppose that there are $n$ signature verifications $J_i ~(i = 1, . . . , n)$ to be scheduled, and each signature has a release time $r_i$, a due time $d_i$, finishing time for verification \textbf {part 1} $C_i $ and the processing requirement of \textbf {part 1} by $p_i$. 
A total schedule is expressed as a set $X$ of signatures such that the total weight $\sum_{i \in X } {w_i}$ is maximal. Besides that, we assume that processing times for all the signature verification \textbf {part 1} are same: $p_i = p$ for all $i$ where $p$ is an arbitrary integer and $r_i + p_i \leq d_i $. Signatures are indexed in a non-decreasing order of due time $d_i$. If the release times $r_i$ are not multiples of $p$ problems, it becomes more complex than problems with unit processing times $p_i = 1$. Such kind of scheduling problem is called:  $  1| r_i;p_j=p | \sum{ w_i U_i} $ where $U$ stands for Unit penalty per late job in the standard scheduling terminology. 

\[
 U_i =
  \begin{cases}
   0 & \text{if } C_i \leq d_i \\
   1       & otherwise
  \end{cases}
\]
\textbf {Example:} $1| r_i;p_j=p$  with given release time $r_i$, due time $d_i$ and $p_i = p =2$

\begin{center}
  \begin{tabular} { l | c | r | c |r | }
    \hline
   ~~~~~~~~~~~~~~~~$i$ ~~& ~1~ & ~2 ~&~3~ &~~ 4~ \\ \hline
   Processing time~$p_i~~$ & ~2~ &~ 2~ & ~2 ~& ~~2~ \\ \hline
Release time ~$r_i$ ~~& ~1~ & ~2~ & ~3~ & ~~8 ~\\ \hline
       Due time ~$d_i$ ~~&~ 3~ &~ 6~ & ~4~ & ~11~ \\ \hline
                Completion time~$C_i$~~ & ~3~ & ~7~ &~ 5~ & ~~10~ \\ \hline

                        Lateness ~$L_i :=(C_i - d_i)$ &~ 0~ &~ 1~ &~ 1~ & ~~$-1$~ \\ \hline

                                Unit penalty ~$U_i$ ~~& ~0~ & ~1 ~&~ 1~ & ~~0~ \\ 
    \hline
  \end{tabular}
\end{center}

\begin{center}
  \begin{tabular}{ l | c  r | }
    \hline
    Max. Completion time~~$C_{max}~~$ &  10& \\ \hline
    Max Lateness ~~$L_{max}~~$ &  1 &  \\ 
    \hline

  \end{tabular}
\end{center}

\graphicspath{{\\C:\Users\Mamun\Desktop\Accepted paper\IEEEtran (1)}} 
\begin {figure}[htb]
\centering
\includegraphics[width=0.5\textwidth]{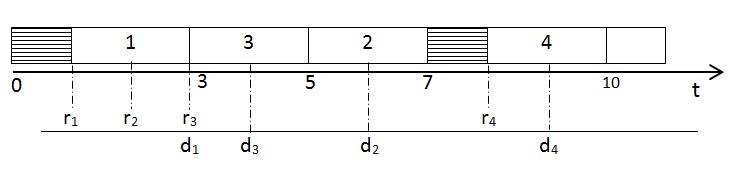}
\caption{Instance for  $1| r_i;p_j=p | $}
\label{fig:  Instance for  $1| r_i;p_j=p | $ }
\end {figure}

A schedule for subset $X$ is feasible if and only if: 
  \begin {enumerate}
 \item All signature verification \textbf {part 1} set $X$ start after or at their release date and are completed before or at their due time, and
\item They  do not overlap in time.
  \end {enumerate}
An optimal schedule will exist if each computation of the signature verification \textbf {part 1} starts at a time belonging to the set.\\
$ T := {r_i + lp~ \vert~ i= 1,....,n; l = 0,...,n-1}$

Let $S$ be an optimal schedule with $i_1, i_2, ...,i_n$ order. It can be transferred to a feasible schedule; for example $i_v$ can be shifted to the left until its release time and due time ($r_v, d_v$) coincide.

For any integer $k\leq n$ and $s,e\in T$ with $s \leq e$. Let $U_k(s,e)$ be the set of verification \textbf {part 1} $i \leq k$ with $s \leq r_i < e$. and $W_k^{*}(s,e) $ is the maximal total weight of a subset of $U_k(s,e)$ with 
 \begin {enumerate}
 \item $S$ is idle before $s + p$ and after $e$
\item The start time of all $i \in T$
  \end {enumerate}

\algsetup{indent = 2em}
\newcommand{\SchdulingVerification} {\ensuremath {\mbox {\sc Signature scheduling  $(1| r_i;p_j=p | \sum{ w_i U_i})$ }}}
\begin{algorithm}[h!]
\caption{$\SchdulingVerification$}\label{alg1:}
 \textbf {INPUT:} $r_i, d_i, p_i$ \\
 \textbf {OUTPUT: Feasible schedule of signatures with maximal total weight ~$W_k(s,e)$.} 
 
\begin {enumerate} 

\item Enumerate the signatures s.t., $d_1\leq d_2\leq ... \leq d_n$;
\item $ \forall s,e \in T$ with $s\leq e : W_0(s,e) := 0$;  
FOR $(k = 1$ TO $n)$\\
~~$ \forall s,e \in T$ with $s\leq e $
    \[
  W_k(s,e) :=
  \begin{cases}
  W_{k-1}(s,e) ~~~ \text{if }  r_k \notin [s,e) \\
        max \{ W_{k-1}(s,e), W_{k}^{\prime}(s,e)\} ~ otherwise
  \end{cases}
\]\\ where
\[
 W_{k}^{\prime}(s,e) := 
    \begin{cases}
  max\{ w_k + W_{k-1}(s,s^{\prime}) + W_{k-1}(s^{\prime},e) \} \\$such that$~~  
  s^{\prime} \in T_i ~$and$\\~ max \{r_k, s+p\} \leq s^{\prime} \leq min \{  d_k,e\} - p;
  \end{cases}
\]
	 
\item Calculate $W_n (s,e)  := ( ~$min$ ~t - p, ~$max$~t  )$ for $t\in T$
\end {enumerate}
\end{algorithm}

\textbf {Theorem	1:} For $k=0,...,n$ and $\forall {s,e} \in T$ such that $s \leq e$ the equality $W_{k}(s,e) ~ =~W_{k}^{*}(s,e)$ holds \cite {Batch Scheduling}.\\ 

For each value of $k, O(n^4)$ values of $W_k$ have to be computed. As $T$ contains $O(n^2)$ elements, overall time complexity of this algorithm is $O(n^7)$. However, when release and due dates of jobs are ordered similarly $([r_i < r_{i+1}] \ [d_i \leq d_{i+1}])$, the same problem is solvable in $O(n^2)$ with a dynamic programming algorithm in \cite {Batch Scheduling3}. In VANET environment, if we do not consider any prioritized message/signature, we can adopt the algorithm in \cite {Batch Scheduling3}.

In \cite {Batch Scheduling4}, authors present a modified tabu search algorithm that also schedules $n$ jobs to a single machine in order to minimize the \textit {maximum lateness} of the jobs. The objective function of the scheduling problem is to minimize~ $L_{max}(\Pi) =$  max~${L_i} $, where $L_i = C_i - d_i$. A total schedule is expressed as a set $ \Pi= \{ \pi(1), \pi(1), ..., \pi(n)\},$ where $\pi(j)$ is the index of the job in position $j$ of the schedule.

In  \cite {Batch Scheduling2} Noy, et al. proposes a single machine throughput maximization real time scheduling algorithm for $n$ jobs where each job is associated with corresponding weight $w_i$ which is the value/revenue of completed job and deadline $d_i$. Let for $k$ batches, they have  $n_j$ as batch size, $F_j$ as family which is associated with processing time. Jobs with same processing time belong to same family and can be executed in the same batch. Their goal was to find a feasible schedule of batching several jobs  of the same type together to maximize weight. This type of problems are referred to as \textit {real time} scheduling problem and maximizing weight referred to as \textit {throughput}.  Authors suggested that if the number of families are fixed, then this type of problems can be solved as dynamic programming. For bounded batch size $b$ they prove and conclude the problem be solved optimally where $F$ is some \textit {constant} value depends on weight of jobs.\\

\textbf {Theorem 2:}  $1|f -$batch$, b, r_j,  F = $const$ | \sum w_j(1- U_j) $ can be solved optimally in $O (n^{{F^2}+3F+3} log ~n)$ steps.\\ 

In our batch scheduling mechanism, we can assume $2$ batches as verification \textbf {part 1} and verification \textbf {part 2} from different family $F_j$, because each batch is responsible for different types of computation as mentioned before. That is why, the batch scheduling problem can be solved optimally with dynamic programming with finite steps. 

We propose \textbf {Algorithm 2} for signature  verification \textbf {part 2}, that is, to find out the optimum value of the number of signatures be batched at a time.

For implementation, we consider $6$ \textit {queues} named  $   \mathbf{M}, \mathbf{B}, \mathbf{K}, \mathbf{Q},   \mathbf{V},   \mathbf{P}  $.  From the \textbf{algorithm 1} we get the optimal schedule of signature verification \textbf {part 1} and hence corresponding values $\mathbf{M_i},~~ \mathbf{B_i},~~ \mathbf{K_i},~~ \mathbf{Q_i}, $~and ~$ \mathbf{\nu_i}$ in the respective named queues. Now we will consider the algorithm for verification \textbf {part 2} which is responsible for pairing computation and incur maximum time among all the computations. Our proposal is running two batches for verification \textbf {part 1} and \textbf {part 2} simultaneously as \textit {two} threads.
\algsetup{indent=2em}
\newcommand{\BatchScheduling}{\ensuremath{\mbox{\sc Scheduling Verification part 2 }}}
\begin{algorithm}[h!]
\caption{$\BatchScheduling$}\label{alg:}
 \textbf {INPUT:}  Setup time $s_b$, Completion time $C_i$, Due time $d_i, 6$ queues $\mathbf{M}, \mathbf{B}, \mathbf{K}, \mathbf{Q},   \mathbf{V},   \mathbf{P}  $ \\
 \textbf {OUTPUT:} (Batch size $ b ~$,  Max. Completion time $C_{max{b}}$, Max. Lateness $L_{max{b}}$.

\begin {itemize} 

\item FOR $(b = 2 ~~$to $n)$
\begin {enumerate} 
\item Pop $i^{th}$ value from $\mathbf{M}, \mathbf{B}, \mathbf{K}, \mathbf{Q},   \mathbf{V},   \mathbf{P}  $ and
 \item Calculate $ \eta =  e( \prod_{i=1}^b \mathbf{B_i},B)  ~~  e(\prod_{i=1}^{b}\mathbf{K_i} ,K_T) ~~\\ e(\prod_{i=1}^{b} \mathbf{Q_i}  ,S)   \prod_{i=t}^{b} \mathbf{\nu_i} $ 
 
 \item Calculate $\mu= \prod_{i=1}^{b} \mathbf{M_i}$ 
 \item Check  $\mu = \eta$. If successful then:
    
     \begin {itemize} 
	\item Push the result of $\eta$ into queue $\mathbf{P}~$  for future use.
   	\item Set operation time of batch size $b$ as $b_t$
	 \item Calculate Maximum Completion time of a batch $C_{max_{b}} := s_b + b_t + C_{max}  $ 
	 \item Calculate Maximum Lateness of a batch $L_{max_{b}} := C_{max_{b}} - d_{i=1} $
           \item Record $(b, ~ C_{max_{b}}, L_{max_{b}})$.
    \end {itemize} 
    
 \item  Else if $\mu \neq \eta$,  report batch error and Exit.

 \item Continue until Queues:$~~   \mathbf{M},~ \mathbf{B},~ \mathbf{K},~ \mathbf{Q},~   \mathbf{V}$ become empty.
\end {enumerate}
\end {itemize}

\end{algorithm}
From the record, one can easily take decision on the optimized batch size $b$ from Maximum completion time $C_{max_b}$ and Maximum Lateness $L_{max_{b}}$ of a batch.
Therefore, the threshold value of batch size $b$ should be chosen carefully to keep the batch verification system remain consistent, flexible and efficient.

 \section {Security and Efficiency}
 
\subsection {Security Analysis.}
The properties of IBGS \cite {signature} allow the vehicles to transmit/receive messages without leaking their identities. The tracing manager TSD can trace a forged identity and revoke the certificate of an anonymous vehicle. To meet the aforementioned security requirement, in \cite {signature} it is shown that IBGS group signature scheme is correct,  anonymous, traceable, and non-frameable. 
The scheme is correct as the message generated by vehicles  will always be accepted by other vehicles and hence guide vehicles for potential improvement of traffic safety and efficiency. If a vehicle is not registered with MVD , revocability, liability and scalability, it cannot create messages even if the cheating vehicle is allowed to access valid messages thorough VANET. An attacker cannot cheat other vehicles even by forging a new valid message or modifying a valid message. This ensures the scheme\rq{s} liability.

The   scheme is non-frameable as no party except the MVD can produce a signature that can be accepted by the verification procedure. This strong security notion guarantees that if a message is accepted as valid,it must have been generated by single registered vehicle and not have been tempered with since it was sent.
The originators of valid messages are anonymous which means an attacker cannot decide the message originator with a probability non-negligibly greater than 1/2. A trusted third party called TSD can trace the anonymous generator of any valid message. No group member or set of colluding members can generate a group signature accepted by the verification procedure which is not linkable to the actual signer. 
\subsection {Efficiency Analysis.}

We compare batch verification with the earlier batch verification mechanism for IBGS \cite {main} as both of them have the same goals and followed the same signature scheme. Our scheme provides the same security and privacy features with a faster verification mechanism. This is due to three reasons: (1) Our new signature size for the batch verification is less than the previous one; (2) We reduce number individual message verifications and batch verifications as well; (3) We propose a runtime batch scheduling algorithm for signature verification and an algorithm to find out the optimum size of the batch. First of all, the previous verification signature was:  $\Upsilon^{\prime} = \Upsilon || (\Gamma_4,\Gamma_6,\Gamma_8) || (\beta_1,\beta_2,\beta_3,\beta_4,\beta_5,\beta_6,\beta_7,\beta_8) || S $ 
where  $\Upsilon = (\Gamma_0,\Gamma_1,\Gamma_2,\Gamma_3,\Gamma_5) || (z_0,z_1,z_2,z_3,Z_1,Z_2,Z_3) ||f||C||v_1||V_2$. Instead, we propose a simplified signature and assume it would be enough for batch verification. We selected the three most complex pairings parameters to pad the original signature, which will give the same functionality as well as less bandwidth and memory consumption. We propose the modified signature as:   $\Upsilon^{\prime} = \Upsilon ||(\beta_4,\beta_6,\beta_8)$. Secondly, we have fewer pairing computations than the previous scheme. Our scheme has only $\textbf {3}$ pairings in final verification while previous scheme had $\textbf {11}$ pairings. Thirdly, we introduce a scheduling algorithm for the incoming signatures, which is very important for VANET environment, where sometimes messages/signatures are prioritized with weights and short deadlines. Finally, we propose an algorithm to explore the best possible batch size $n$ for any specific environment by parallelizing the independently computationable components to set maximum signatures to be processed without any idle time between them. We believe that our algorithms can be implemented for other batch verification algorithms as well to find out the optimum number of signatures to be processed at a time in batch. However, as we used Small Exponent test for batch verification, we accept the probability of invalid signature $2^{l_b}$, where $l_b$ is the security level \cite {mihir_batch}.

\section {Conclusion}
In group signature mechanisms usually large numbers of signatures need to be verified. Designing a batch verification mechanism partially addresses this problem, but batch verifications need  to be optimized as much as possible. In this paper we have presented an extremely efficient batch verification system from a IBGS group signature scheme for VANET environments. We have further presented a batch scheduling algorithm to accelerate batch verification. Our scheme not only provides the desired level of security requirements, but also is efficient in storage and computation.  We believe it can be implemented in any ad hoc network with minimum resource constraints, especially in MANET environments. Our batch scheduling environment can be applied to any batch verification systems available. In future, we would like to evaluate the result on a large scale VANET testbed with varying different scheduling algorithms.

\end{document}